\documentclass[runningheads]{llncs}
\usepackage{graphicx}
\usepackage{subcaption}
\usepackage{booktabs}       
\usepackage{float}
\floatstyle{plaintop}
\restylefloat{table}

\begin{document}
\title{Sound Model Factory: An Integrated System Architecture for Generative Audio Modelling\thanks{This research was supported by 
a Singapore MOE Tier 2 grant, “Learning Generative Recurrent Neural Networks”, and by an NVIDIA Corporation Academic Programs GPU grant.}}
\titlerunning{Sound Model Factory}
%
\author{Lonce Wyse\orcidID{0000-0002-9200-1048} \and
Purnima Kamath\orcidID{0000-0003-0351-6574} \and
Chitralekha Gupta\orcidID{0000-0003-1350-9095}}
%
\authorrunning{Wyse, Kamath, and Gupta}
%
\institute{National University of Singapore\\
\email{\{lonce.wyse, purnima.kamath, chitralekha\}@nus.edu.sg}}
%
\maketitle              
\begin{abstract}
We introduce a new system for data-driven audio sound model design built around two different neural network architectures, a Generative Adversarial Network(GAN) and a Recurrent Neural Network (RNN), that takes advantage of the unique characteristics of each to achieve the system objectives that neither is capable of addressing alone. The objective of the system is to generate interactively controllable sound models given (a) a range of sounds the model should be able to synthesize, and (b) a specification of the parametric controls for navigating that space of sounds. The range of sounds is defined by a dataset provided by the designer, while the means of navigation is defined by a combination of data labels and the selection of a sub-manifold from the latent space learned by the GAN. Our proposed system takes advantage of the rich latent space of a GAN that consists of sounds that fill out the spaces “between” real data-like sounds. This augmented data from the GAN is then used to train an RNN for its ability to respond immediately and continuously to parameter changes and to generate audio over unlimited periods of time. Furthermore, we develop a self-organizing map technique for ``smoothing" the latent space of GAN that results in perceptually smooth interpolation between audio timbres. We validate this process through user studies. The system contributes advances to the state of the art for generative sound model design that include system configuration and components for improving interpolation and the expansion of audio modeling capabilities beyond musical pitch and percussive instrument sounds into the more complex space of audio textures.

\end{abstract}
\section{Background and Motivation} 
Sound modeling for the purpose of synthesis is still mostly a manual programming process. It is labour intensive, costly, and the implementation is done by specialists other than end users of the systems. The overarching objective of this work is to provide a means for instrument and sound designers to create  interactive sound generators by merely providing data samples from a sound space and specifications for how they want to navigate that space.

We use the term ``sound model" rather than ``synthesizer" in part to emphasize the process of their creation, but also because synthesizers are typically designed to produce as wide a range of sounds as possible. Sound models are more targeted and have two essential components: a constrained range of sounds they are capable of producing, and an \emph{expressive} means of navigating that space via parametric control. 

A sound model for our purposes also meets several ``playability" criteria. One is that parameters can be changed at arbitrary times. We define the Parameter Response Time (PRT) as the number of time steps for sound generation to adapt to changes in the input or conditioning parameters. Playability requires perceptually negligible PRT. A playable sound model should also be capable of generating sound for an unlimited period of time.  It further requires an understandable mapping between changing control parameters and their effect on sound. We operationalize this notion to mean (a) that different parameters pertain to different perceptual dimensions, and (b) that their affect on perceptual changes in sound are approximately linear or at least smooth.  

The system presented here is constructed to address all sound and not limited to musical or pitched sounds. In terms of both sound and control strategies, we aspire to universality not in the individual sound models, but in the capabilities of the machines used to create them. Manual programming of sound models is already ``constrained only by our imagination", but has practical time and effort limitations. In this paper, we are working toward the development of a data-driven system for creating arbitrary sound models. We refer to the system as a Sound Model Factory (SMF).

    \subsection{Previous work} 
Neural network and deep learning approaches in particular have been making steady progress towards data-driven sound modeling. Esling et al. \cite{esling2018generative} creates latent timbre spaces with a Variational Autoencoder (VAE) regularized to match perceptual distance metrics for pitched musical instruments, and then traverses the space through potentially novel sounds using direct latent vector paths or audio descriptors (e.g. spectral centroid). The system exhibits 'playability' characteristics such as being responsive to parameter changes since the VAE is trained on single spectral frames of data, though such short data frames would not be capable of capturing complex textures with longer time dependencies.  

WaveNet \cite{van2016wavenet} was a seminal contribution with its multi-level dilated convolutional autoregressive approach and extremely high-quality synthesis, and it is possible to train waveNet conditionally for parametric interaction. Engel et al. \cite{engel2017neural} put WaveNet to work as an autoencoder that learns latent representations for the temporal structure of musical notes. For the 4-second musical notes comprising the NSynth database introduced in the same paper, a 125-step, 16 channel temporal embedding is learned, each vector representing a 32ms window of the note. This code sequence is then used as conditional control for a second WaveNet serving as a decoder. Frame rate ``morphing" between instruments works well, but the temporal representations requires interpolating between two vectors that are themselves changing at each time step during the decoding process. The model is  also capable of generating unseen note sequences and playing notes of longer duration than those it was trained on. The study was restricted to pitched musical notes, so its applicability to unpitched sounds is not known.  

In many ways, GANSynth \cite{engel2019gaNSynth}, also from the Google group, has held the mantle of 'state-of-the-art' in generative musical instrument modeling since it was published. It pioneered the use of a two-channel time-frequency representation of audio training data where one channel is the familiar magnitude spectrogram, and the other is the derivative of phase which is commonly referred to as the ``instantaneous frequency." This ``IF" representation can be inverted to time domain audio using inverse Fourier transforms without the need for a process such as the iterative Griffin-Lim \cite{GriffinLim84} procedure. GANSynth produces excellent quality reconstruction on the NSynth musical instrument notes dataset. 

Where the autoregressive WaveNet produces a single value of an audio waveform at each point in time, GANSynth trains to produce an extended waveform (e.g. a note) for each latent parameter vector input sacrificing high-resolution parametric control for parallelizable computation.  Using a method developed for Conditional GANs \cite{mirza2014conditional} Engle et al. \cite{engel2019gaNSynth}, showed that by training GANSynth conditionally, pitch can be disentangled from the timbral qualities of the sound, and then controlled during generation without having to search the latent space. For conditioning, they used a one-hot representation of pitch to augment the latent vector input. 

The expressivity of GANSynth was demonstrated with a performance of Bach's Prelude Suite No.~1\footnote{https://magenta.tensorflow.org/gansynth} while morphing through dozens of different identifiable as well as novel instrument sounds over the course of the piece, however the GAN's fixed output duration limits the resolution of the morphing to the duration of the notes.   

The IF representation works well with pitched sounds comprised of harmonic components well-separated in frequency, but struggles with unpitched sounds\cite{gupta2021signal}. Noisier sounds have been addressed by others with different representations.  Nistal et al. \cite{nistal2020drumgan} used a 2-channel real and imaginary spectrogram with a GANSynth architecture, conditioning on a set of high-level timbral descriptions (e.g. brightness roughness) each represented as floating point values, to produce convincing drum sounds. Antognini et al. \cite{antognini2019audio} used a CNN architecture inspired by the style-transfer approach using Gram matrices pioneered by \cite{gatys2015texture} and developed for audio \cite{audio2016style,bin2019applying,grinstein2018audio,antognini2019audio}. Data are represented as magnitude spectrograms with time as the single dimension and frequency bins as channels and then inverted with the Griffin-Lim \cite{GriffinLim84} algorithm. They also developed novel loss terms so that the style-transfer algorithm would capture rhythmic time scale patterns. Caracalla and Roebel  \cite{caracalla2020sound} developed a method using the Gram matrix representations but optimizing the time domain signal directly so that no inverse spectrogram operation is necessary significantly improving the perceptual quality of impacts and noisy signals. 

These strategies enable the generation of non-pitched sounds and textures, but are limited, like GANSynth, to generating sounds of a predetermined length (typically on the order of 1-4 seconds) given a parameter or sound as input. They are not  continuously responsive to parameter changes despite the continuity in the latent space.  Conditioning with one-hot representations, as GANSynth does for pitch, also impedes morphing in that dimension, producing between-note transitions resembling cross fades rather than pitch glides as the one-hot vectors are interpolated.

The SMF system described herein makes contributions extending the state-of-the-art GAN system for generative audio modeling with the capacity to produce models that can generate pitched or noisy sounds, a method for linearizing the data space of the GAN, and the ability to generate continuous audio morphing between sounds learned by the GAN with immediate parameter response times in both conditioned and unconditioned dimensions.      
      
\section{Architecture}
\subsection{System Overview}  
\begin{figure*}[h]
      \centering
      \includegraphics[width=1\linewidth]{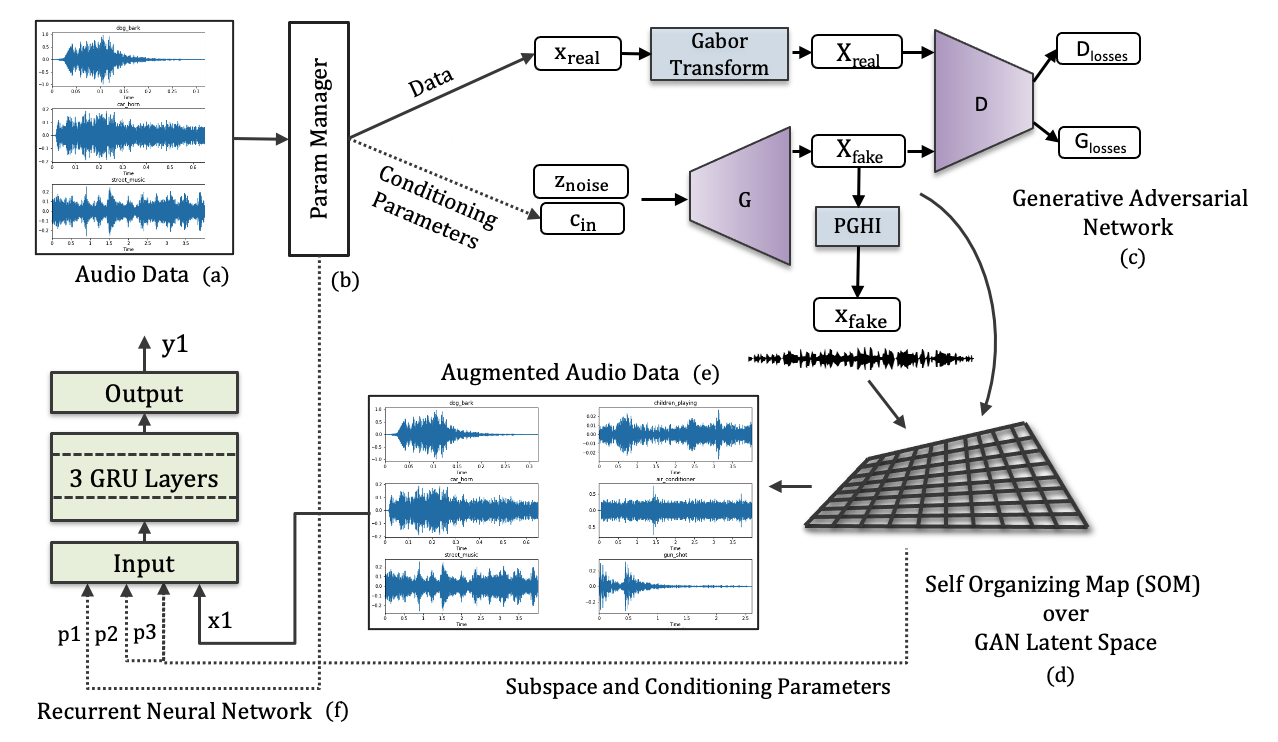}
      \caption{A schematic of the sound model factory is shown in this figure. (a) Curated datasets and optionally parameters from meta-data (b) are used to train the GAN (c). A subspace is chosen from the resulting latent space which is then adaptively adjusted using a Self-Organizing Map (d). The synthetic data (e) is generated from the resulting mesh and paired with the low-dimensional mesh parameters to conditionally train the RNN (f).}
      \label{fig:system_schema}
    \end{figure*}

The SMF is comprised of multiple components with a GAN, a self organizing map (SOM), and an RNN forming the backbone of the flow as depicted in  Figure \ref{fig:system_schema}. To put more familiar functional labels on the components, the GAN functions as an ``interpolator" for creating a continuous space of sounds that fills out regions of parameter spaces between sounds used in the training set. The SOM is a ``smoother" for making the rate of change more consistent as the parameter space is navigated during generation, and the RNN is the ``performer" as the end product of the SMF work flow providing the playability features necessary for expressive interaction. 

The model design process starts with the specification of the dataset (Fig.\ref{fig:system_schema}a). The ParamManager (Fig.\ref{fig:system_schema}b) represents any system for reading and writing data and metadata including parameters, any of which can be chosen to be used as conditioning extensions to the latent noise vectors provided as input for the GAN (Fig.\ref{fig:system_schema}c). 

The next step is the selection of a subspace from the high-dimensional latent space of the GAN that will define the range of sounds of the model being constructed, and the number of dimensions that the sound model will offer for parametric control. We currently select 4 points to define a 2D subspace, but the process generalizes to more dimensions. For example, having trained on the NSynth data base, one might choose two clarinet-like sounds, one at the low end of the pitch scale, the other at the top, and likewise two points for a trumpet-like sound spanning a pitch scale. A 2D mesh (Fig.\ref{fig:system_schema}d)  (of any desired resolution) with corners at the four points, defines the points in latent space that will be used to generate the sounds (Fig.\ref{fig:system_schema}e) that, along with the two parameters used to index them, will be used to train the RNN (Fig.\ref{fig:system_schema}f). A Self Organizing Map (described in more detail below) also smooths mesh interpolations to create more intuitive parametric control. The sound model output of the SMF is the trained RNN. 

   \subsection{System Components}\label{components}

    We used two datasets in the development of the SMF; one consisting of pitched sounds, the other of noisy sounds with no clear pitch. 
    
	We use a subset of the NSynth dataset \cite{engel2017neural} that consists of ~300,000 single-note examples played by more than 1,000 different instruments. It contains various labels for pitch, velocity, instrument type, and more, although, for this particular work, we only make use of the pitch information as the conditional parameter. We further narrow the set to acoustic instrument types \textit{reed} and \textit{brass}. Like \cite{nistal2021comparing}, we trimmed the audio samples from 4 to 1 seconds, choosing the central 1 second to remove the attack and decay transitions and any silence. We considered samples with a MIDI pitch range from 64 to 76, which finally yields a subset of approximately 6,164 audio files with 3,041 of brass type and 3,123 of reed type.
    
The BOReilly dataset\footnote{https://animatedsound.com/datasets/} consists of 6,580 one-second samples of synthetic textures generated by 8 different synthesizers drawn from source material used by sonic artist Brian O'Reilly\footnote{https://vimeo.com/dendriform}. The only criteria in this loosely curated collection was that the sounds be noisy and unpitched (in contrast to the NSynth set) and of stable amplitude throughout (comparable to our NSynth subset).
    
 We use the progressively growing Wasserstein GAN architecture developed for GANSynth \cite{engel2019gaNSynth}  consisting of a generator G and a discriminator D, where the input to G is a random vector $z$ with 128 components from a spherical Gaussian distribution along with an optional one-hot conditional vector $c_{in}$. Training is divided into 5 stages, where the resolution of the output progressively increases at each stage by adding a new layer to the existing stack. This gradual and progressive blending in of the new layers ensures minimal perturbation effects as well as stable training of the GAN, as first proposed by  \cite{karras2017progressive}. We train separate GAN models for different datasets, each trained for 1.2M iterations, with 200k iterations in each of the first three stages on batches of 12, and 300k in the last two stages on batches of 8, that requires up to 1.5 days on an NVIDIA Tesla V100-SXM2-32GB. The Wasserstein GAN minimizes an approximation of the Earth Mover (EM) distance, that has been shown to manage issues with previous GANs such as mode collapse \cite{arjovsky2017towards,arjovsky2017wasserstein}. The discriminator $D$ estimates the Wasserstein distance between the real and the generated distributions. 

Similar to \cite{engel2019gaNSynth}, for the NSynth dataset, we sometimes train conditionally on pitch with the goal of achieving independent control of pitch and timbre. To do that, a one-hot representation of musical pitch is appended to the latent vector. Unconditional GAN training (no learning of control parameters) was used for the BOReilly dataset or when combining the BOReilly and NSynth datasets into a single sound model.
 
For audio data representation, rather than the 2-channel IF used for GANSynth, we use a single channel, 2-dimensional log magnitude spectrum and the Phase Gradient Heap Integration (PGHI) method \cite{pruuvsa2017noniterative} for a non-iterative method to reconstruct the time signal. Marafioti et al. \cite{marafioti2019adversarial} found that the PGHI algorithm worked well on synthetic spectra produced by GANs. \cite{gupta2021signal} showed that training the GANSynth architecture using the PGHI representation and inversion produces results equivalent to the IF representation for pitched sounds, and significantly better audio quality for wideband, non-pitched or fast changing signals such as pops, and chirps. Time-frequency representations of 16kHz sampled audio are computed using an FFT size of 512 with a hop size of 128 samples (i.e.~75\% overlap between consecutive frames). 

The RNN used in the SMF was originally developed in Wyse \cite{wyse2018real} for modeling NSynth data and is comprised of 3 gated recurrent unit (GRU) layers \cite{cho2014learning} with 256 hidden nodes each. A fully connected embedding layer takes the floating point audio and conditioning parameters to the GRU layer input. A softmax layer at the output produces a probability distribution across 256 values interpreted as mu-law encoded audio sample values predicting the next sample in time. RNNs are trained on the sound samples generated by the GAN latents sampled on a 21x21 grid, and conditioned on the parameters that index the grid. This stage was run on an NVIDIA 3090 on 256 sample length sequences in batches of 128 for 100K iterations. 

\section{Connecting the GAN and the RNN}        
The size of the  GAN latent space (in our case, a fairly typical 128 dimensions) is too large to serve as a parametric interface for effective human control of musical performance. This motivates the dimension reduction accomplished by choosing points that define a subspace.  There are various ways of ``inverting" a GAN to find parameters corresponding to specific data space examples \cite{xia2021gan}.  However, in practice we have found that generating a hundred or two random vectors does a reasonable job at generating sounds similar to much of the real data used for training. The designer auditions random samples and chooses a set which circumscribes the range of sounds the final synthesis model will be able to generate, as well as how they will be parametrically navigated. 

After the choice of latent vectors/sounds is made, they are used as corners defining a subspace that is then sampled with a mesh pattern. Each low-dimensional mesh point indexes a point in the much higher-dimensional GAN latent space.  The GAN is used to generate sounds for all points on the mesh. Then the pairing of each mesh point and the resulting GAN-generated sound can be used to train the RNN. Original and adapted grids generated in this way  are available for auditioning online for the Trumpinet and BOReilly sets\footnote{https://animatedsound.com/evomusart2022/\#grids}.

\subsection{Parameter Linearization}

GANs model the structure of the data space using an explicit distribution of latent vectors that allow the learned probability density function to be sampled. The space tends to be smooth in that nearby points in latent space yield nearby points in the output audio data space\cite{goodfellow2016nips,mao2019tunagan,jahanian2019steerability} giving rise to the GAN's remarkable ability to interpolate, or ``morph" between data points. However, there is no guarantee that the perceptual space is linear with respect to the latent space. That is, a given delta in the latent parameter values can cause very different changes in the perception of the sound depending on the location in latent space. In practice, we have found that for latent values that correspond to sounds similar to those in the real dataset, parameter changes tend to have relatively small effects, where as at some points in the latent space between regions of real-data like sounds, there is a kind of boundary where small parameter changes result in large perceptual changes.  

To address the issue of consistent interpolation which is central to sound model design, we introduce a  Self Organizing Map (SOM) \cite{kohonen2012self} that sits in the SMF system between the learned space of the GAN and the training of the RNN (Figure \ref{fig:system_schema}). The latent vectors function as the SOM ``weights" associated with each point on a 2D mesh, and the reduction of a measure of the perceptual difference between sound generated by neighboring  latent vectors on the grid serves as the function driving the adaptation so that differences between sounds generated by neighboring mesh points become perceptually more uniformly distributed. The mesh point update equation is 
\begin{equation} \label{eq:SOM}
 g_{i,j}  \mathrel{+}= \delta   \sum_{m,n\in N} g_{m,n} \left(\sum_{p,q} |S_{p,q}^{i,j} -  S_{p,q}^{m,n} |^2\right)^\frac{1}{2}
\end{equation}
where $N$ is the 8-node neighborhood (horizontal, vertical, and diagonal) of $g_{i,j}$, $\delta$ is the step size, $S^{i,j}$ is the spectrogram generated by the GAN by the latent vector associated with mesh node $i,j$, and ${p,q}$ indexes the time and frequency points of the spectrograms. The corners of the grid are always ``pinned" to the initial latent vector chosen to define the sub-manifold, and for some experiments, we restrict movement of mesh points lying on the lines connecting corners to remain on those lines.

The adaptation of the latent vectors on the mesh is driven by the L2 distance between the 2D spectrogram images as a proxy for perceptual difference. Although audio perceptual distance is notoriously difficulty to quantify in general, sound corresponding to neighboring latent vectors are already organized by spectral proximity. We find that linearizing distances locally in this way yields significant improvements to the overall perceptual smoothness of the parameter space, and we validate this claim with user studies reported below.

\begin{figure*}[h]
      \centering
      \includegraphics[width=1\linewidth]{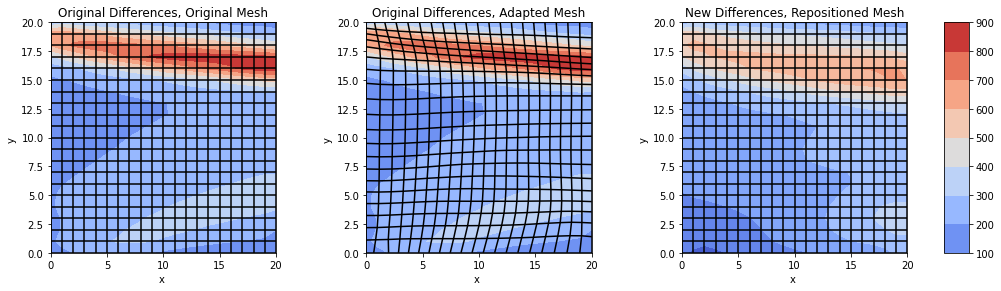}
      \caption{The SOM adapting to reduce differences between mesh neighbors of the function value. The upper two mesh corners correspond to a clarinet and a trumpet sound playing the same pitch. The bottom two corners are complex BOReilly textures. A border-like structure of sudden change divides the two regions.  The contour plot shows average spectrogram differences with it 8 neighbors at each grid location. (a) Initial mesh (b) adapted mesh (with edges pinned) over the original differences contour (c) the smoothed differences at their new mesh positions.}
      \label{fig:SOM adaptation}
    \end{figure*}

We can visualize the mesh adaptation using the 2D grid space (Figure \ref{fig:SOM adaptation}). The contour plot represents the average difference in the spectrograms produced by one grid point and its eight neighbors. The effect of the adaptation can be seen by stretching the grid points back to their nominal index value locations which stretches the underlying space with it. In Figure \ref{fig:SOM adaptation}, the linearization of the perceptual space is seen in the regularization of the distance between contour lines on the contour plot and the reduction of the maximum difference. 

\section{Evaluation}
As a system, we want to evaluate the capabilities compared to others that address the same or similar goals - the production of sound models that have minimal PRTs, can generate non-repetitive sounds for arbitrary lengths of time, and that can interpolate smoothly between parameter values. We also want to understand the kinds and quality of sound the models created by the system are capable of generating, particularly in terms of complexity. 
        
\subsection{Human Evaluation adaptively smoothed latent space}\label{humaneval}
Our listening tests were designed to test for perceptual linearity between two points in the GAN's latent space by asking participants to listen to two sound clips, one sampled from the adapted mesh space (pinned corners or pinned corners with constrained edges) and another sampled without this adaptation. The sound clips were created by first selecting two points in the GAN latent space, selecting 20 equidistant samples between those points and joining those samples to create one clip. The space between those two points was then adapted using our SOM technique and sampled again to create a second clip. These two clips formed one pair in a comparison trial with each sound lasting for 13-15 seconds. The mesh adaptation for the BOReilly sounds used pinned corners whereas for NSynth sounds used pinned corners with constrained edges. Our listening experiments were conducted on Amazon's Mechanical Turk (AMT) with user consent, are anonymized, and were approved by our university department's ethics review committee.   
        
We recruited 40 participants on AMT and given that these participants may not be audio or sound experts, our experiment design used simple language and pairs of icons (see Table \ref{table:listeningtest-results-table}) to convey our intent and to gather meaningful responses from the test. The participants were required to listen to both sounds before answering the questions. No submissions from participants were rejected during these experiments. The participants were paid  \$0.10 per task which took less than 90 seconds on average. 
        
We identified two types of non-linearity in sound interpolations. One type is when the interpolation takes a ``detour" (e.g. passing through a flute sound during a transition from a clarinet to a trumpet, or passing through a pitch not between the pitch of the endpoint sounds) and another type when step sizes in the interpolation parameter results in the perception of irregular size steps in the quality of the sound. To probe for directness in transitions of sounds from one end point to the other, we asked participants to associate a ``direct" icon  or a ``detour" icon with each pair of sounds. To test for irregular step sizes in the sampled GAN data space, the participants did the same with an ``Even Steps" icon and a ``Uneven Steps" icon. The mutually exclusive association of each pair of ions with the sounds could be made with one mouse click and drag \footnote{https://animatedsound.com/evomusart2022/ui/amt-instructions-template.html}.  Responses from 520 comparison trials for NSynth sounds and 840 comparison trials for BOReilly sounds were collected. The trials were loaded in a random sequence to AMT and its results are tabulated in Table \ref{table:listeningtest-results-table}.
        
\begin{table}[!htbp]
            \begin{minipage}{0.6\linewidth}
                \centering
                \begin{tabular}{llll}
                    \toprule
                                        & Sound samples   & NSynth    & BOReilly\\
                    \midrule
                    Direct vs.          & Mesh adapted  & \textbf{57.50\%}   & \textbf{61.07\%}\\
                    \cmidrule(r){2-4}
                    Detour              & GAN Only      & 42.50\%   & 38.92\%\\
                    \midrule
                    Even steps vs.      & Mesh adapted  & \textbf{59.42\%}   & \textbf{60.47\%}\\
                    \cmidrule(r){2-4}
                    Uneven Steps        & GAN only      & 40.58\%   & 39.52\%\\
                                        
                    \bottomrule
              \end{tabular}
            \end{minipage}\hfill
        	\begin{minipage}{0.35\linewidth}
        		\centering
        		\includegraphics[width=40mm]{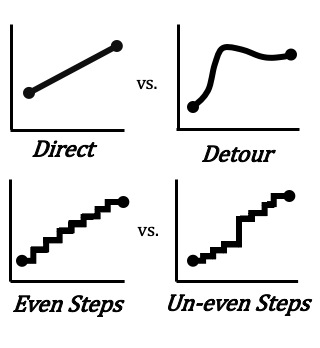}
    	    \end{minipage}
    	    \caption{Human evaluation of sound navigation through the SOM-adapted space vs. the unadapted GAN grid (left).  Mechanical Turk participants used the icons (right) to visually associate perceptual differences in the sounds in comparison trials. Icons represent the perception of a sound sequence as it transitions from its starting point to its end (Direct vs. Detour) as well as the perceptual measure of even or uneven jumps the sounds makes in its transition (Even vs. Uneven steps).} 
    	    \label{table:listeningtest-results-table} 
    	 \end{table}

\subsection{Parameter Response Time} 
\begin{figure}[h]
\centering
\begin{subfigure}{0.3\textwidth}
\includegraphics[width=\textwidth]{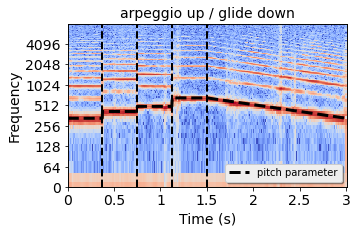} 
\caption{Arpeggio and glide}
\label{fig:responsiveness_pitch_subim1}
\end{subfigure}
\begin{subfigure}{0.45\textwidth}
\includegraphics[width=\textwidth]{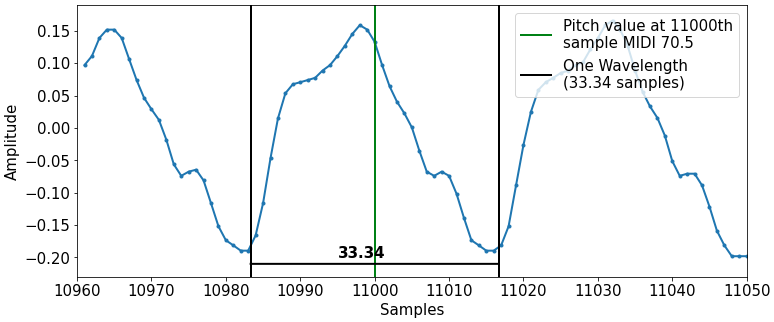}
\caption{Wave period during pitch glide}
\label{fig:glide_analysis_midi70}
\end{subfigure}

\begin{subfigure}{0.8\textwidth}
\includegraphics[width=\textwidth]{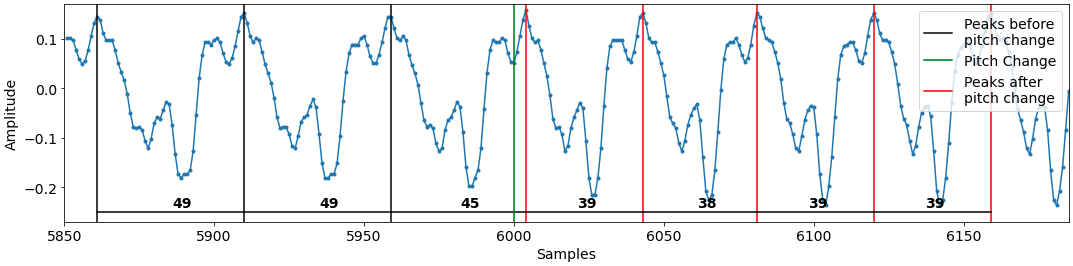}
\caption{Wave period at pitch step change}
\label{fig:step_analysis}
\end{subfigure}
\caption{(a) Parameter changes and spectrogram of RNN output for (a) an arpeggio over an octave (midi number 64, 68, 71, 76) followed by a one octave glide. The PRT is more precisely quantified in (b) for a descending one-octave glide (MIDI note 76-64), the period of the waveform centered at the 11000 sample mark matches the gliding pitch  parameter of MIDI note 70.5 (34.35 sample wavelength) exactly at that point, and (c) for pitch step where we identify peaks before and after the pitch change (green line) to show that the wave period changes immediately after the pitch steps from MIDI 64 (48.56 sample wavelength) to 68 (38.53 sample wavelength). }
\label{fig:prt_analysis}
\end{figure}

In this section, we demonstrate the ability of the SMF system to generate sound models that meet the playability requirements identified in the beginning of this paper. The RNN was trained using only same steady-state pitch parameters used for training the GAN on NSynth data, but RNN parameters can be arbitrarily manipulated during generation. Figure \ref{fig:responsiveness_pitch_subim1} shows a one-octave stepped arpeggio followed by a descending one-octave glide. We measure the PRT manually by verifying whether the wave period at a particular instance corresponds to the intended pitch at that instant. We find that the parameter response is immediate for both steps and glides as shown in Figure \ref{fig:prt_analysis}.
         
\subsection{Sound Quality Evaluation based on Audio Classification}
First we quantitatively test whether the pitch and timbre qualities of a GAN-generated audio are similar to RNN at the end of the multistage flow of the proposed SMF for the same parameter settings. GAN generated audio  has been shown to be of high perceptual quality (\cite{engel2019gaNSynth,nistal2021comparing}), and we hope to preserve the quality through the various stages.   

Since the latent space of GAN consists of timbre of the two instrument classes as well as the timbres in-between the two classes, a class label cannot be assigned to every point in this latent space, as the in-between timbres cannot be labelled with a class. Therefore, the standard method of inception score computation (\cite{salimans2016improved}) will not be applicable in this case. 

For quantitative validation, we train an audio classification network (\cite{palanisamy2020rethinking}) with the brass and reed instrument classes from NSynth dataset. This subset has 56-76 MIDI pitch values with velocity 127, with a total of 2,225 audio files, split into 80\% training and 20\% validation. The audio classification network consists of a ResNet model initialized with ImageNet pretrained weights. This scheme of training has achieved state-of-the-art audio classification performance when fine-tuned with standard audio classification datasets, as shown by \cite{palanisamy2020rethinking}. We trained this network to classify the two instrument classes by fine-tuning it on our Nsynth subset data. The validation accuracy of this model is 100\%.

\begin{figure}[h]
    \centering
    \begin{subfigure}{0.61\textwidth}
    \includegraphics[width=\textwidth]{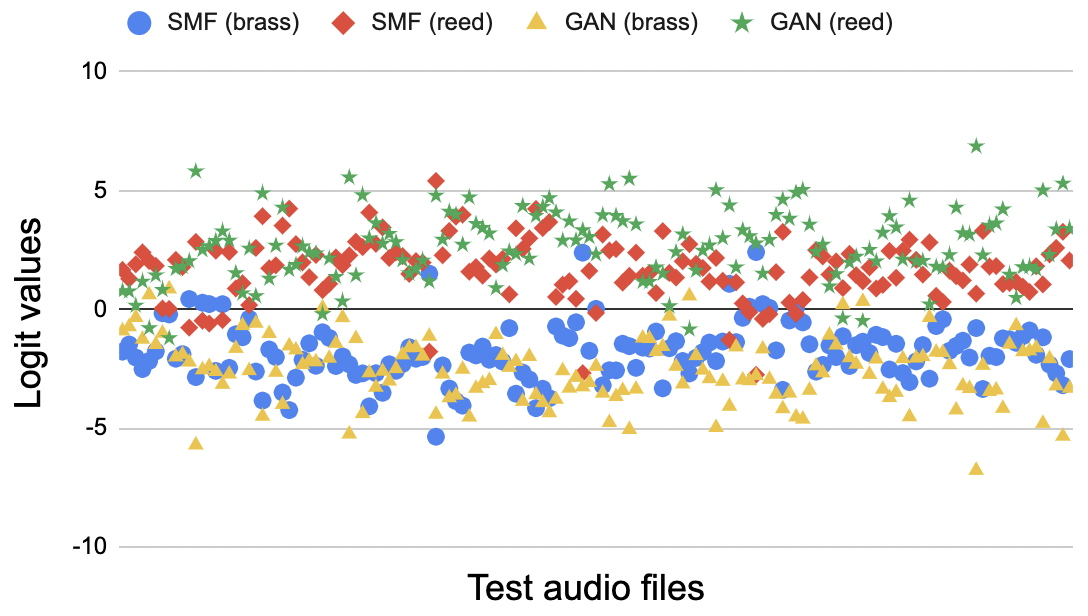}
    \caption{}
    \label{fig:logitvalues}
    \end{subfigure}
    \begin{subfigure}{0.38\textwidth}
    \includegraphics[width=\textwidth]{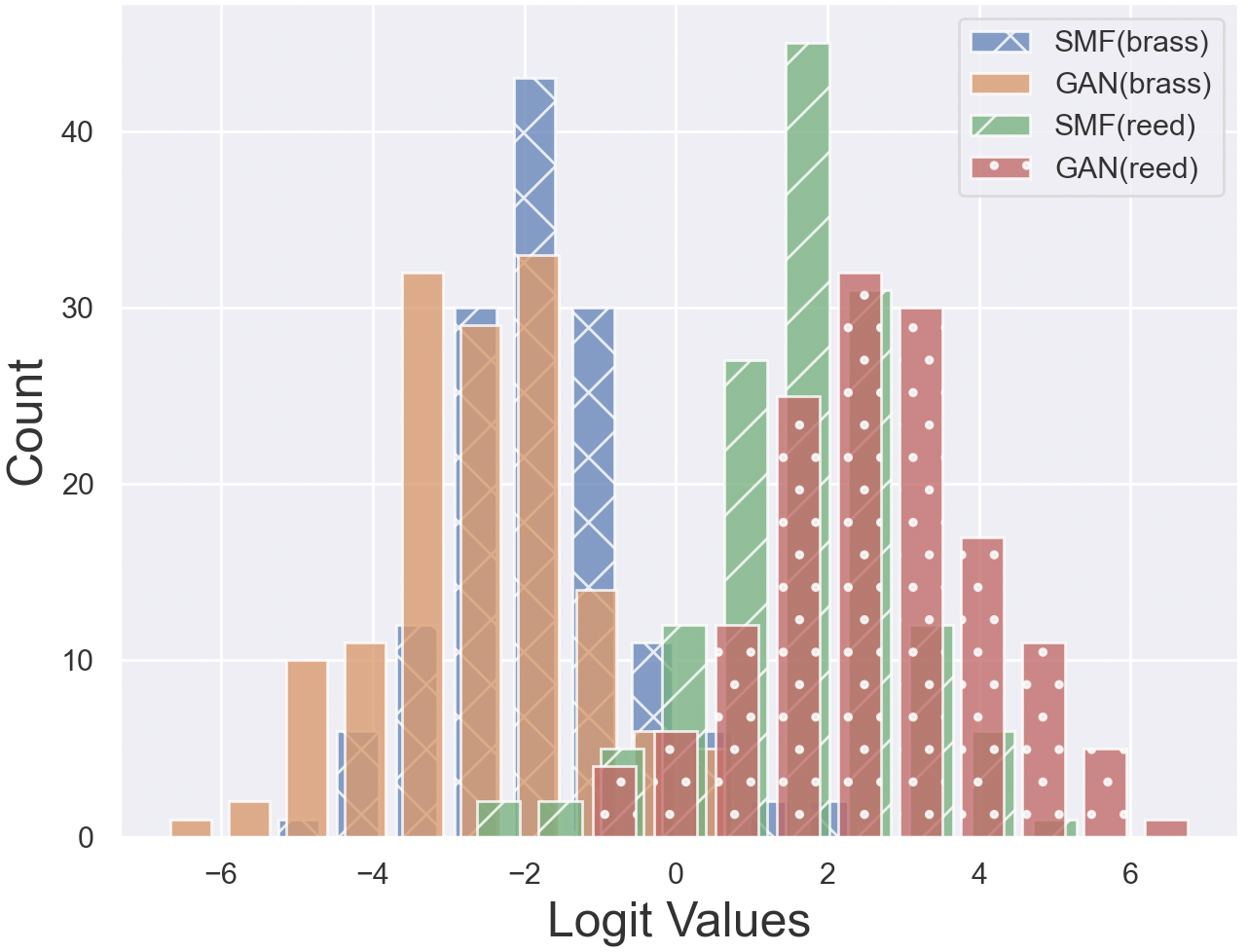}
    \caption{}
    \label{fig:logitvalueshist}
    \end{subfigure}
    \caption{Logit values from the penultimate output layer of the audio classifier for the two sets of 143 test audio files, one set from the GAN, and the other from the proposed SMF model (GAN+RNN). The two logit values are depicted as *(brass) and *(reed). (a) The raw logit values for all test audio files, and (b) the histogram of the logit values.}
\end{figure}

In this experiment, we were interested in observing whether a set of audio files generated from a latent space of the GAN have the same distribution of logit values (values at the output layer before the class labels are assigned) in the audio classifier, as the set of audio files generated from the proposed SMF (i.e.~GAN+RNN) system for the same parameter settings. Parameter settings, in this case, means the pitch and timbre indices from the GAN-generated latent space. We used 143 such pairs of audio files. The two logit values corresponding to the two audio classes (brass and reed) for each of these files, i.e.~(SMF(brass),SMF(reed)) and (GAN(brass),GAN(reed)), are presented in Figure \ref{fig:logitvalues}. The histogram of these logit values are shown in Figure \ref{fig:logitvalueshist}. Qualitatively, we observe the distribution of logit 0 values of GAN and SMF audio files are similar, and the distribution of logit 1 values of the two types of audio files are similar. Moreover, 127 out of 143 pairs of audio files get classified into the same instrument class, i.e. 88.8\% of the test pairs are classified into the same class. This analysis shows that the audio generated from the proposed SMF architecture are similar to that generated from the GAN.

\subsection{Continuous interpolation of pitch and timbre}
\begin{figure}[h]
\centering
\begin{subfigure}{0.3\textwidth}
\includegraphics[width=\textwidth]{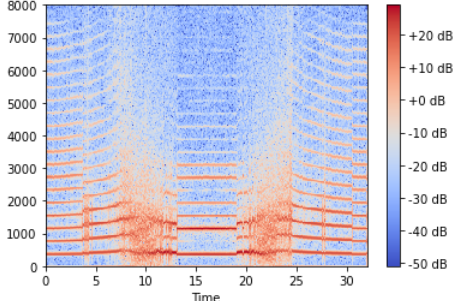} 
\caption{}
\label{fig:timbreglide_endpointonly}
\end{subfigure}
\begin{subfigure}{0.31\textwidth}
\includegraphics[width=\textwidth]{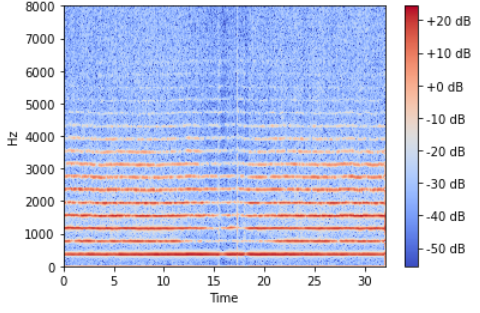}
\caption{}
\label{fig:timbreglide_gangridtrained}
\end{subfigure}
\begin{subfigure}{0.31\textwidth}
\includegraphics[width=\textwidth]{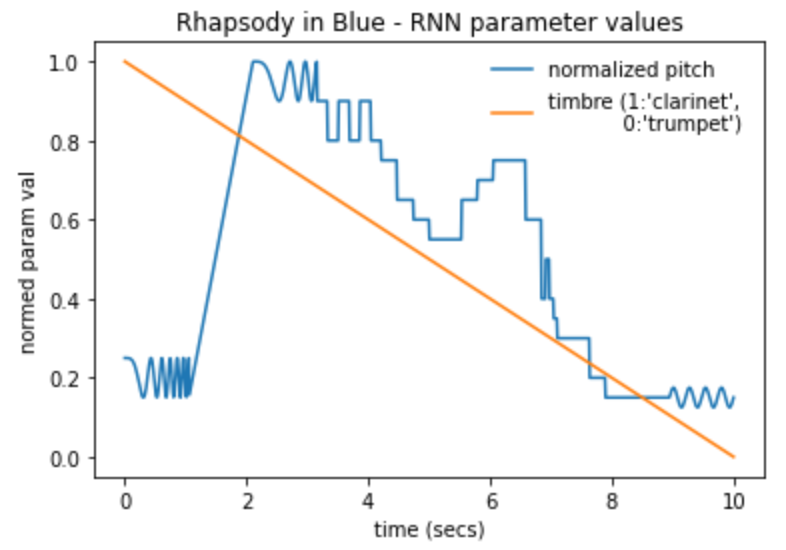}
\caption{}
\end{subfigure}
\caption[Caption for LOF]{Spectrograms of a steady-pitch trumpet-clarinet-trumpet glide with an RNN (a) trained at endpoints only, (b) trained with intermediate synthetic data from GAN.  (c) Parameter sequence for a ``reinterpreted" opening segment of George Gershwin's Rhapsody in Blue with glides, various note lengths, vibrato, and a continuous clarinet-to-trumpet instrument morph (audition online\footnotemark).}
\label{fig:timbreglide}
\end{figure}

GANs are structurally unable to continuously interpolate parameters for audio because time is a dimension of the data representation. One-hot representations such as used for pitch in GANSynth produce audible artifacts at ``in between" values. Wyse \cite{wyse2018real} showed that an RNN can interpolate between widely spaced conditioned pitch values. However, we have found the RNN is less capable of interpolating between timbres as can be seen in Fig. \ref{fig:timbreglide}a. In contrast, the GAN can synthesize convincing timbres perceptually between training examples. When synthetic GAN data are used to train the RNN, the result is the best of both worlds - convincing in-between timbres that can be smoothly and continuously interpolated (Fig. \ref{fig:timbreglide}b). A musical example illustrates the combined capabilities with fast and slow pitch changes including vibrato and a continuous morph between clarinet and trumpet over a rendering of the opening segment of George Gershwin's Rhapsody in Blue (Fig. \ref{fig:timbreglide}c).

\footnotetext{Audition at https://animatedsound.com/evomusart2022/\#pitch-timbre}

\section{Conclusion}\label{conclusion}
    
A few noteworthy limitations are left for future work. For example, by selecting a subspace of the GAN latent space, the SMF makes a trade-off between low dimensional control over a constrained sound space of the RNN and the hundreds of parameters and diversity of the GAN sound space. More control over that balance would benefit sound model design.
    
The linearization of the perceptual sound space with respect to parameters needs further exploration. Although the SOM (\ref{eq:SOM}) improved the perceived smoothness of transitions, they are still neither perceptually or objectively uniform. The neighborhood used to compute updates to the grid could be expanded, or paths between the anchor points could even depart from the 2D manifold in a search for smoother transitions similar to the way Esling et al. used  sound descriptors for directing navigation through latent space. 
    
Although the PGHI representation and inversion provides high quality resynthesis of non-pitched sounds, the current SMF is limited in the complexity of the temporal structure of audio textures it can capture. For example, previous synthesis models capable of learning to generate randomly spaced clicks conditioned on rate have needed lots of data to capture the temporal variability at any given rate parameter. However, GANs generate one fixed length sample for any given latent parameter vector. If the GAN is to provide the necessary variety of sound patterns for training a downstream model with this kind of complexity, we would have to identify a whole region of the GAN latent space to map to a single parameter value for training the RNN, not just a single point. 
    
The innovations introduced at the system level for the SMF come from a) the use of the single channel and PGHI  representation and reconstruction with the GANSynth architecture, b) the linearization of a slice of the latent space of the GAN using SOMs, and (c) the exploitation of the GAN parameter/sound pairing to train an RNN. The SMF contributes novel audio capabilities by combining immediate parameter response times, smooth interpolation capabilities of the GAN's conditioned and unconditioned parameters, ``infinite" duration sound synthesis, and the ability to generate a wider variety of audio timbres more complex than pitched instrument sounds.

%
%
%
\bibliographystyle{splncs04}
\bibliography{Wyse}

\begin{thebibliography}{10}
\providecommand{\url}[1]{\texttt{#1}}
\providecommand{\urlprefix}{URL }
\providecommand{\doi}[1]{https://doi.org/#1}

\bibitem{antognini2019audio}
Antognini, J.M., Hoffman, M., Weiss, R.J.: Audio texture synthesis with random
  neural networks: Improving diversity and quality. In: ICASSP 2019-2019 IEEE
  International Conference on Acoustics, Speech and Signal Processing (ICASSP).
  pp. 3587--3591. IEEE (2019)

\bibitem{arjovsky2017towards}
Arjovsky, M., Bottou, L.: Towards principled methods for training generative
  adversarial networks. arXiv preprint arXiv:1701.04862  (2017)

\bibitem{arjovsky2017wasserstein}
Arjovsky, M., Chintala, S., Bottou, L.: Wasserstein generative adversarial
  networks. In: International conference on machine learning. pp. 214--223.
  PMLR (2017)

\bibitem{caracalla2020sound}
Caracalla, H., Roebel, A.: Sound texture synthesis using ri spectrograms. In:
  ICASSP 2020-2020 IEEE International Conference on Acoustics, Speech and
  Signal Processing (ICASSP). pp. 416--420. IEEE (2020)

\bibitem{cho2014learning}
Cho, K., Van~Merri{\"e}nboer, B., Gulcehre, C., Bahdanau, D., Bougares, F.,
  Schwenk, H., Bengio, Y.: Learning phrase representations using rnn
  encoder-decoder for statistical machine translation. arXiv preprint
  arXiv:1406.1078  (2014)

\bibitem{engel2019gaNSynth}
Engel, J., Agrawal, K.K., Chen, S., Gulrajani, I., Donahue, C., Roberts, A.:
  Gansynth: Adversarial neural audio synthesis. arXiv preprint arXiv:1902.08710
   (2019)

\bibitem{engel2017neural}
Engel, J., Resnick, C., Roberts, A., Dieleman, S., Norouzi, M., Eck, D.,
  Simonyan, K.: Neural audio synthesis of musical notes with wavenet
  autoencoders. In: International Conference on Machine Learning. pp.
  1068--1077. PMLR (2017)

\bibitem{esling2018generative}
Esling, P., Bitton, A., et~al.: Generative timbre spaces: regularizing
  variational auto-encoders with perceptual metrics. arXiv preprint
  arXiv:1805.08501  (2018)

\bibitem{gatys2015texture}
Gatys, L., Ecker, A.S., Bethge, M.: Texture synthesis using convolutional
  neural networks. In: Advances in Neural Information Processing Systems. pp.
  262--270 (2015)

\bibitem{goodfellow2016nips}
Goodfellow, I.: Nips 2016 tutorial: Generative adversarial networks. arXiv
  preprint arXiv:1701.00160  (2016)

\bibitem{GriffinLim84}
Griffin, D., Lim, J.: Signal estimation from modified shorttime fourier
  transform. IEEE Transactions on Audio, Speech and Language Processing
  \textbf{32}(2),  236–243 (1984)

\bibitem{grinstein2018audio}
Grinstein, E., Duong, N.Q., Ozerov, A., P{\'e}rez, P.: Audio style transfer.
  In: 2018 IEEE International Conference on Acoustics, Speech and Signal
  Processing (ICASSP). pp. 586--590. IEEE (2018)

\bibitem{gupta2021signal}
Gupta, C., Kamath, P., Wyse, L.: Signal representations for synthesizing audio
  textures with generative adversarial networks. arXiv preprint
  arXiv:2103.07390  (2021)

\bibitem{bin2019applying}
Huzaifah, M., Wyse, L.: Applying visual domain style transfer and texture
  synthesis techniques to audio: insights and challenges. Neural Computing and
  Applications  \textbf{32}(4),  1051--1065 (2020)

\bibitem{jahanian2019steerability}
Jahanian, A., Chai, L., Isola, P.: On the" steerability" of generative
  adversarial networks. arXiv preprint arXiv:1907.07171  (2019)

\bibitem{karras2017progressive}
Karras, T., Aila, T., Laine, S., Lehtinen, J.: Progressive growing of gans for
  improved quality, stability, and variation. arXiv preprint arXiv:1710.10196
  (2017)

\bibitem{kohonen2012self}
Kohonen, T.: Self-organization and associative memory, vol.~8. Springer Science
  \& Business Media (2012)

\bibitem{mao2019tunagan}
Mao, W., Lou, B., Yuan, J.: Tunagan: Interpretable gan for smart editing. arXiv
  preprint arXiv:1908.06163  (2019)

\bibitem{marafioti2019adversarial}
Marafioti, A., Perraudin, N., Holighaus, N., Majdak, P.: Adversarial generation
  of time-frequency features with application in audio synthesis. In:
  International Conference on Machine Learning. pp. 4352--4362. PMLR (2019)

\bibitem{mirza2014conditional}
Mirza, M., Osindero, S.: Conditional generative adversarial nets. arXiv
  preprint arXiv:1411.1784  (2014)

\bibitem{nistal2021comparing}
Nistal, J., Lattner, S., Richard, G.: Comparing representations for audio
  synthesis using generative adversarial networks. In: 2020 28th European
  Signal Processing Conference (EUSIPCO). pp. 161--165. IEEE (2021)

\bibitem{nistal2020drumgan}
Nistal, J., Lattner, S., Richard, G.: Drumgan: Synthesis of drum sounds with
  timbral feature conditioning using generative adversarial networks. arXiv
  preprint arXiv:2008.12073  (2020)

\bibitem{van2016wavenet}
van~den Oord, A., Dieleman, S., Zen, H., Simonyan, K., Vinyals, O., Graves, A.,
  Kalchbrenner, N., Senior, A., Kavukcuoglu, K.: Wavenet: A generative model
  for raw audio. arXiv preprint arXiv:1609.03499  (2016)

\bibitem{palanisamy2020rethinking}
Palanisamy, K., Singhania, D., Yao, A.: Rethinking cnn models for audio
  classification. arXiv preprint arXiv:2007.11154  (2020)

\bibitem{pruuvsa2017noniterative}
Pr\r{u}{\v{s}}a, Z., Balazs, P., S{\o}ndergaard, P.L.: A noniterative method
  for reconstruction of phase from stft magnitude. IEEE/ACM Transactions on
  Audio, Speech, and Language Processing  \textbf{25}(5),  1154--1164 (2017)

\bibitem{salimans2016improved}
Salimans, T., Goodfellow, I., Zaremba, W., Cheung, V., Radford, A., Chen, X.:
  Improved techniques for training gans. Advances in neural information
  processing systems  \textbf{29},  2234--2242 (2016)

\bibitem{audio2016style}
Ulyanov, D., Lebedev, V.: Audio texture synthesis and style transfer.
  https://dmitryulyanov.github.io/audio-texture-synthesis-and-style-transfer/
  (2016), accessed: 2019-07-10

\bibitem{wyse2018real}
Wyse, L.: Real-valued parametric conditioning of an rnn for interactive sound
  synthesis. In: Proceedings of the Proceedings of the 6th International
  Workshop on Musical Metacreation. ACM Conference on Computational Creativity,
  Salamanca, Spain (2018)

\bibitem{xia2021gan}
Xia, W., Zhang, Y., Yang, Y., Xue, J.H., Zhou, B., Yang, M.H.: Gan inversion: A
  survey. arXiv preprint arXiv:2101.05278  (2021)

\end{thebibliography}
%
\end{document}